\documentstyle[prd,aps,floats,tabularx,epsfig]{revtex}

\begin{document}
\draft
\preprint{\
\begin{tabular}{rr}
&
\end{tabular}
}
\twocolumn[\hsize\textwidth\columnwidth\hsize\csname@twocolumnfalse\endcsname
\title{The shape of a small universe: signatures in the cosmic
microwave background}
\author{R.~Bowen \&
  P.~G.~Ferreira}
\address{Astrophysics, University of Oxford,
NAPL, Keble Road, Oxford OX1 3RH, UK}

\maketitle

\begin{abstract}
We consider the most general parametrization of flat topologically compact
universes, complementing the work of Scannapieco, Levin and Silk to
include non-trivial shapes. We find that modifications in shape
of the fundamental domain will lead to distinct signatures in
the anisotropy of the cosmic microwave radiation. We make a preliminary
assessment of the effect on three statistics: the angular power spectrum, the
distribution of identified ``circles'' on the surface of last scattering
and the correlation function of antipodal points. 
\end{abstract}
\date{\today}
\pacs{PACS Numbers : 98.80.Cq, 98.70.Vc, 98.80.Hw}
]
\renewcommand{\thefootnote}{\arabic{footnote}} \setcounter{footnote}{0}
\noindent
With the dramatic improvement of cosmological observations in the past few
years, efforts are underway to precisely characterize the nature and
structure of space-time. One property which has generated much interest
is the possibility that we live on a manifold which has a compact, non-trivial
topology \cite{topo}. 
A multitude of tests and observables have been proposed which would
allow us to pin down the topological class of the universe we live in.
Under the assumption that we live in a smooth Friedman-Robertson-Walker 
universe with perturbations set up adiabatically in the early universe, 
measurements of the Cosmic Microwave Background anisotropy seem to indicate 
that our universe is flat \cite{3exp}. This greatly reduces the set of models
one needs to look at and simplifies the boundary conditions of the
wave functions one needs to consider. Scannapieco, Levin and Silk
\cite{sls}
have identified complete sets of functions for each topology for the
case where the identification vectors are orthogonal. 

In the context of Kaluza-Klein theories, it has recently been pointed out
in \cite{dienes,mafi} that non-orthogonality of the identification
vectors will lead to interesting effects in the mass differences of 
Kaluza-Klein states. It was shown in \cite{dienes} that the mass gap of the 
graviton will become dependent on what he calls the shape parameter, i.e.
the angle between identification vectors. While he considered the case
of a compact 2-dimensional space rather than the standard 3-dimensional
space we live in, we shall see what happens if we apply such a
transformation to a compactified version of our 3-dimensional space
in the case of no extra dimensions.

Without loss of generality let us consider the simplest, flat, non trivial
topology- the hypertorus in the classification of \cite{ellis}. This will allow us to outline the formalism which can be
used in the general case. Consider a 3-manifold with coordinates
 ${\vec \xi}$=($\xi_1,\xi_2,\xi_3$) on which any function, $f(\xi_1,\xi_2,\xi_3)$ satisfies the following
property:
\begin{eqnarray}
f(\xi_1+2\pi,\xi_2,\xi_3)=f(\xi_1,\xi_2+2\pi,\xi_3)\nonumber \\=f(\xi_1,\xi_2,\xi_3+2\pi)=f(\xi_1,\xi_2,\xi_3)
\end{eqnarray}
We can think of this manifold as tiled by cubes with sides of length $2\pi$.
Alternatively we can think of it as simply consisting of a cube with
opposite faces identified; for convenience let us call it
a {\it fundamental domain}. We can now rescale each direction in such a way
as to have different identification scales in each direction by defining
a matrix ${\bf L}$:
\begin{equation}
{\bf L}={\rm diag}(R_1,R_2,R_3)
\end{equation}
and a new set of coordinates: ${\vec \chi}={\bf L}\cdot {\vec \xi}$.
It will be useful to divide this transformation into two: one that
is simply the product of the identity by some overall scale factor, 
$R$ and another transformation which is diagonal with unit determinant.
The first transformation corresponds to an overall volume transformation,
modifying the volume of the fundamental domain by a factor of $R^3$.
The second transformation induces anisotropic scale differences in
the sides of the fundamental domain (changing it into a paralliliped).
A coordinate invariant definition orders the eigenvalues of
this second transformation, 
$\lambda_1>\lambda_2>\lambda_3$ and defines $\sigma_1=\lambda_2/\lambda_1$
and $\sigma_2=\lambda_3/\lambda_1$.

We now wish to consider an additional linear transform that changes
the directions of the identification vectors. Naturally we are not
interested in overall rotations of all vectors, given that ultimately
any observable we construct should be rotationally invariant. Consider
a set of 3 vectors: ${\vec s_1}$, ${\vec s_2}$, ${\vec s_3}$. Given that
we wish to factor out overall rotations, we can align ${\vec s_1}$ with
$\xi_1$, have ${\vec s_2}$ lie on the ($\xi_1$,$\xi_2$) plane and ${\vec s_3}$
will have an arbitrary direction. We can define a shape transformation
matrix, ${\bf S}$, such that its three columns are the three vectors, 
${\vec s_i}$. A convenient parametrization is the following:
\begin{eqnarray}
{\bf S}=\omega
\left( \begin{array}{ccc}
1 & \cos{\alpha_1} & \cos{\alpha_2} \\
0 & \sin{\alpha_1} & \sin{\alpha_2}\cos{\alpha_3} \\
0 & 0 & \sin{\alpha_2}\sin{\alpha_3}
\end{array} \right)
\end{eqnarray}
Note that the the prefactor, $\omega=(\sin\alpha_1\sin\alpha_2\sin\alpha_3)^{-1/3}$,
 ensures that $\det{\bf S}=1$.
This parametrization is such that 
${\vec s_1}\cdot{\vec s_2}=\omega^2\cos{\alpha_1}$,
${\vec s_1}\cdot{\vec s_3}=\omega^2\cos{\alpha_2}$ and 
${\vec s_2}\cdot{\vec s_3}=\omega^2(\cos{\alpha_1}\cos{\alpha_2}+
\sin{\alpha_1}\sin{\alpha_2}\cos{\alpha_3})$. The geometrical interpretation
of this transformation is simple: we are changing the angles of
the corner of our fundamental domain while preserving its volume. 
We can now consider a new set
of coordinates ${\vec x}={\bf S}{\bf L}\cdot {\vec \xi}$.
The identification vectors are not orthogonal anymore. 

Note that we are allowed to perform these transformations without 
jeopardizing the suitability  of the manifold as a cosmological
model. We start off with a locally homogeneous and isotropic
space time with non-trivial identifications which are permitted.
We then simply apply a linear transformation which preserves
the homogeneity and isotropy of the manifold as well as
differentiability.
Indeed if we
consider a generalization to a flat manifold in $N$ dimensions with
analogous topological constraints, we are allowed to perform any
non-singular linear transformation and the resulting manifold will
still have the same differentiability class. Such a transformation
will be characterized in terms of $N$ rescaling parameters
and $N(N-1)/2$ shape parameters. Furthermore, these transformations
clearly cannot change the topology class of the manifold. In the
example we are considering, the topology remains that of a
hypertorus and the geometry remains the same: zero curvature.

The plane
wave solutions to the Laplaces equations will be modified.
A complete set of orthogonal solutions is given by 
$\psi_{\vec n}({\vec \xi})=\exp(-i{\vec n}\cdot{\vec \xi})$
with ${\vec n}$=$2\pi$($n_1$,$n_2$,$n_3$). This can be mapped directly
onto a complete set of orthogonal solutions on the transformed space:
$\phi_{\vec k}({\vec x})\equiv\exp(-i{\vec k}\cdot{\vec x})=
\psi_{\vec n}({\vec \xi})$ where the last equality implies
${\vec k}={\bf L}^{-1}{\bf S}^{-1}{\bf n}$. As above the
wavenumbers are quantized, but the effect of ${\bf S}$ is to
mix up the quantization constraints through the shape angles.
The explicit form of the wavenumber for this topology is
\begin{eqnarray}
\frac{\omega R_1k_1}{2\pi}&=&n_1-n_2\cot\alpha_1+n_3
(\cot\alpha_1 \cot\alpha_3-\frac{\cot\alpha_2}{\sin\alpha_3})\nonumber \\
\frac{\omega R_2k_2}{2\pi}&=&n_2\frac{1}{\sin\alpha_1}-n_3\frac{\cot\alpha_3}{\sin\alpha_1} 
\nonumber \\
\frac{\omega R_3k_3}{2\pi}&=&n_3\frac{1}{\sin\alpha_2\sin\alpha_3}.
\end{eqnarray}
In this note we wish to focus on the effect of shape. We will
restrict ourselves to $R_1=R_2=R_3=R$ and two simple, one paramater
families of shape
transformations:
\begin{description}
\item[A:] We fix $\alpha_2=\alpha_3=\pi/2$ and vary $\alpha_1=\alpha\in[0,\pi/2]$.
This essentially corresponds to collapsing one of the identification
vectors of the fundamental domain on the plane defined by the other two.
\item[B:] We fix ${\hat s}_1\cdot{\hat s}_2
={\hat s}_2\cdot{\hat s}_3={\hat s}_3\cdot{\hat s}_1=\cos\alpha$ and
vary $\alpha\in[0,\pi/2]$.
\end{description}

\begin{figure}[!t]
\begin{center}
\includegraphics[angle=0,totalheight=7cm,width=8.cm]{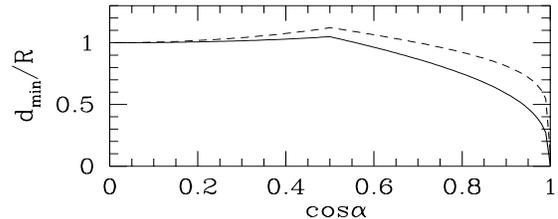}
\end{center}
\vskip -1in
\caption{\label{fig1}The minimum traversible distance for
the two familys of shape transformations we are considering,
as a function of the cosine of the angle. The solid (dashed) line
corresponds to type A (B).}
\end{figure}

An hypothesis was put
forward in \cite{sls} that the minimum traversable distance of
the space defines the scale at which observable effects appear
 in statistics of the CMB such as the angular power spectrum
or the geometry and frequency of ghost images. Although
our shape transformation has unit determinant, the minimum traversable
distance ($d_{min}$) may  be greater or less than the one in the
cubic domain.  For the two families we above we find that
\begin{eqnarray}
d_{min}=R\omega\left\{\begin{array}{ll}
1 & \mbox{for $\cos\alpha\in[0,0.5]$} \\
\\
\sqrt{2(1-\cos\alpha)} & \mbox{for $\cos\alpha\in[0.5,1]$}
\end{array}\right.
\end{eqnarray}
and in Fig \ref{fig1} we plot this quantity
for the two families we are considering. We find that, indeed, $d_{min}$ does
vary as a function of shape parameter. Although there is a
slight increase for $\cos\alpha<0.5$, both parametrizations
A and B lead to a dramatic decrease in $d_{min}$ for $\cos\alpha\sim1$. 
The question we now wish to address is, how will this be refelected
in some of the statistics that have been proposed to quantify 
topology from the CMB.

Let us first focus on the angular power 
spectrum of the CMB. In a flat universe, the anisotropy in the CMB,
$\Delta T({\hat  \theta})$ (where ${\hat \theta}$ 
is the unit vector pointing in
a given direction in the sky), is dominated by the Sachs-Wolfe term, i.e. 
\begin{equation}
\Delta T({\hat \theta})=-\frac{1}{3}\Phi(\eta_*,{\hat \theta}d_*)
\end{equation}
where $\Phi(\eta,{\bf x})$ is the gravitational potential, $\eta_*$
is the time of last scattering, and $d_*$ is the comoving distance to
the surface of last scattering. We will be working in Fourier space where
$f({\bf x})=\sum_{\vec k}f({\vec k})\exp(-i{\vec k}\cdot{\vec x})$. Assuming 
statistical homogeneity and isotropy we have that 
$\langle\Phi^*(\eta_0,{\vec k})\Phi^*(\eta_0,{\vec k}')\rangle=
{\cal P}(k)\delta_{{\vec k},{\vec k}'}$. Defining 
$a_{\ell m}=\int d{\hat \theta}T({\hat  \theta})Y^*_{\ell m}({\hat  \theta})$
we have that
\begin{eqnarray}
\langle a^*_{\ell m}a_{\ell' m'}\rangle=(4\pi)^2i^{l'-l}\sum_{\vec k}
{\cal P}(k)j_\ell(kd_*)j_{\ell'}(kd_*) && \nonumber \\
\times Y_{\ell m}({\hat  k})
Y^*_{\ell' m'}({\hat  k}) &&
\end{eqnarray}
In general, for arbitrary topology the covariance matrix of the $a_{\ell m}$
will not be diagonal. The fact that there are preferred orientations
in space (defined by the identification vectors) will induce correlations
in the $a_{\ell m}$ between different $\ell$ and $m$ which depend on
the orientation of the $Y_{\ell m}({\hat \theta})$ basis. In this
letter, as in \cite{sls}, we wish to look at the overall impact
on a rotationally invariant measure of the angular power spectrum.
We shall consider 
\begin{eqnarray}
C_\ell\equiv\frac{1}{2\ell+1}\sum_{m=-\ell}^{\ell}|a_{\ell m}|^2=
(4\pi)\sum_{\vec k} {\cal P}(k)j^2_\ell(kd_*)
\end{eqnarray}
For the infinite volume case this converges to the standard
angular power spectrum.

\begin{figure}[!t]
\begin{center}
\includegraphics[angle=0,totalheight=7cm,width=8.cm]{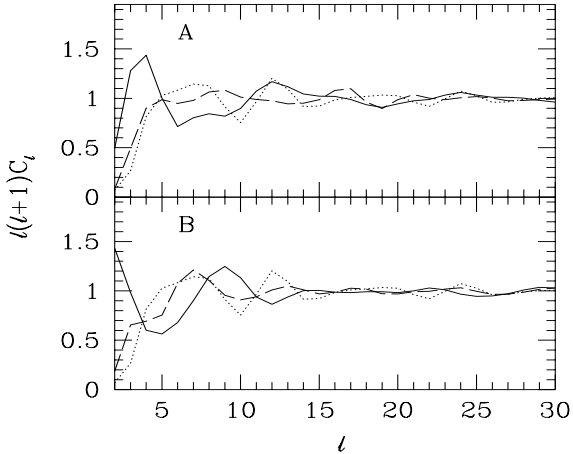}
\end{center}
\caption{\label{fig2}The angular power spectrum for a selection of three
values of the shape parameter for the two families  of transformations, A and
B. The angle $\alpha$ takes values $\pi/2$ (dotted), $\pi/4$ (long-dashed) and
$\pi/8$ (solid).}
\end{figure}

In Figure \ref{fig2} we plot angular power spectra
for $\alpha=\pi/2$, $\pi/4$, and $\pi/8$. We are considering
a universe with an identification scale $R=d_*$. For the
cubic domain the features identified by \cite{sls} are manifest, i.e.
one finds a supression of power for low $\ell$ and a set of
oscillatory features out to large $\ell$. Changing the value of
$\alpha$ will shift the oscillatory features around but can have
a curious effect at the lowest multipoles: the solid lines
(corresponding to $\alpha=\pi/8$) show an increase in power due
to the fact that the shape transformation can lead to the
regeneration of small wavenumber modes in the integral. Indeed
we find this to be a general trend.

All real space statistics rely, to some extent,
 on the repetition of patterns on
the surface of last scattering. The surface of last scattering
will intersect points in space which are identified. This leads
to perfect correlations between certain points (or pixels) of
a map of the CMB. The more points which are identified, the
more significant a detection of topology will be. We can attempt
to quantify the effect of shape by estimating how many points
(or structure, such as for example circles) will be identified
for a given value of $\alpha$ in either of the transformations,
A and B. Two points ${\vec x}$ and ${\vec x'}$ lie on the surface
of last scattering if they satisfy $|x|^2=|x'|^2=d^2_*$ and can be
identified with one another if there exist a trio of integers,
$(n_1,n_2,n_3)$ such that ${\vec x'}={\vec x}+n_1{\vec s}_1
+n_2{\vec s}_2+n_2{\vec s}_2$. Combining these two sets of
equations one finds that a given trio of integers
satisfies this condition if
\begin{eqnarray}
(n_1^2+n_2^2+n_3^2
+2n_1n_2{\hat s}_1\cdot{\hat s}_2
+2n_2n_3{\hat s}_2\cdot{\hat s}_3 \\ \nonumber
+2n_3n_1{\hat s}_3\cdot{\hat s}_1)^{1/2}/
{[{\hat s}_1\cdot({\hat s}_2\times{\hat s}_3)]^{1/3}}<\frac{2d_*}{R}.
\end{eqnarray}
For example, for an orthogonal trio of ${\hat s}_i$ and $R=2d_*$ one has 
3 pairs of identified structures. This condition for transformation
A is:
\begin{eqnarray}
(n_1^2+n_2^2+n_3^2
+2n_1n_2\cos\alpha)/(\sin\alpha)^{1/3}<\frac{2d_*}{R}
\end{eqnarray}
and for transformation B is:
\begin{eqnarray}
[n_1^2+n_2^2+n_3^2
+2\cos\alpha(n_1n_2+n_2n_3+n_3n_1)]^{1/2}/\nonumber \\
{(1-\cos\alpha)\sqrt(1+2\cos\alpha)]^{1/3}}<\frac{2d_*}{R}.
\end{eqnarray}
As $\cos\alpha\rightarrow1$, more  roots
are possible, leading to stronger correlations between points on the
surface of last scattering.

There is in fact a simple way of understanding how the number of roots
(or matched structures) increases with $\cos\alpha$. Let us first consider
transformation A. The direction of the minimum traversable distance
is given (with $\cos\alpha>0.5$) by ${\vec s}_1-{\vec s}_2$. As we increase
$\cos\alpha$, we decrease $d_{min}$ so that the number of identified
structures will be given by the integer part of $2d_*/d_{min}$. Given that
$\lim_{\cos\alpha\rightarrow1}d_{min}=0$, there will be more
matched structures with increasing $\cos\alpha$. For transformation
B, the reasoning is the same except there are now three possible directions:
${\vec s}_1-{\vec s}_2$,${\vec s}_2-{\vec s}_3$ and ${\vec s}_3-{\vec s}_1$.
We then have that the number of matched structures is given by the
integer part of $6d_*/d_{min}$.

Let us now focus on two real space statistics that have been advocated
for identifying topology in the CMB: matching circles in the sky
by \cite{cs} and the correlation function of antipodal points of
\cite{lsgsb}. In \cite{cs} Cornish, Spergel, and Starkman have 
proposed matching
the circles which arise from the intersection of the surface of
last scattering with the identified boundaries of the fundamental domain.
One way of thinking about this is to consider a given identification
vector ${\vec s}$. One can then construct two perpendicular planes at
each end and find the lines of their intersection with the sphere of
radius $d_*$ and center at the midpoint of the vector. For example,
for a cubic domain of size $R<2d_*$ one expects at least 3 pairs of
circles with centers on the 3 axes of identification. Pairs of
circles will intersect (with each other) if $R<\sqrt{2}d_*$, and each
circle will subtend an angle $\theta_{0}=
\frac{\pi}{2}-\arcsin\frac{R}{2d_*}$
on the sky. For a cubic domain only pairs of big circles will intersect.
Let us now consider a fundamental domain with a non-trivial shape
parameter. The identification vectors will not be orthogonal and
it will be possible to have pairs of small circles that intersect.
The condition is that $\alpha<\theta_{0}$. This is then a
signature which could distinguish between cubic and non-cubic
domains. More generally, if one identifies more than one pair of circles
and the vectors  joining their centers are not perpendicular,
then there is evidence for non-trivial shape. Naturally one would
have to distinguish it from the hexagonal fundamental domains,
where the identification vectors are at $60^{\rm o}$.

\begin{figure}[!t]
\begin{center}
\includegraphics[angle=0,totalheight=7cm,width=8.cm]{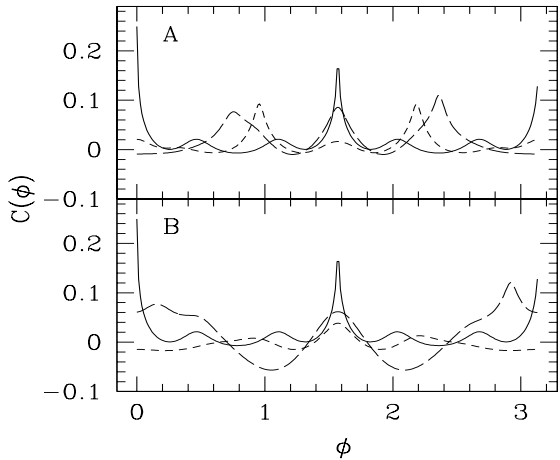}
\end{center}
\caption{\label{fig3}A one dimensional slice of the antipodal correlation
function for a selection of three
values of the shape parameter for the two families  of transformations, A and
B. The angle $\alpha$ takes values $\pi/2$ (solid), $\pi/4$ (short-dash) and
$\pi/8$ (long-dash).}
\end{figure}

Levin {\it et al.} in \cite{lsgsb} 
have suggested that a good statistic for identifying
non-trivial topology is the correlation of antipodal points on
the surface of last scattering:
\begin{eqnarray}
C_A({\hat \theta})\equiv\langle \Delta T(-{\hat \theta})\Delta T({\hat \theta})
\rangle\propto\sum_{\bf k}{\cal P}(k)\cos(2d_*{\bf k}\cdot{\hat \theta}).
\end{eqnarray}
The patterns which appear will reflect the geometry of the fundamental
domains. Again one might expect that a non-cubic domain may lead
to a different signature; for example it has been shown in \cite{lsgsb}
that a hexagonal fundamental domain will have a very different
signature than a cubic one. As an illustration of this statistic for
non-trivial shapes we have calculated $C_A({\hat \theta})$ for the one
dimensional slice ${\hat \theta}=(\cos\phi,\sin\phi,0)$. 
We can cleary see the differences in Fig. \ref{fig3} where we plot
$C_A(\phi)$ for the two families of shape transformations.

In this report we have taken a first look at how the shape of
the fundamental domain of a flat, topologically non-trivial
universe might affect a few different statistics of the CMB.
There clearly seem to be observable signatures which will
constrain the values of the shape parameters. We will now finish
with a few comments. To begin with, it is clear that when it
is claimed that one can use the cosmic microwave background
to constrain the topology of the universe, it is in fact meant
that one is constraining certain geometric properties which
arise when the universe is topologically compact. For example, if
the identification scales are infinite, then the shape transformations
are irrelevant. When choosing the topology of the universe
(from the six possibilities discussed in \cite{sls}), one must
also include the 6 parameters which describe the geometry
and scale of the fundamental domain.
One can then proceed to apply the variety of statistics that
have been advocated. It is possible, however, to incorporate
the parameters describing the fundamental domain
into the standard likelihood methods
which have been used to constrain cosmological parameters.
The standard parameters 
$p_i$ for Gaussian models, should be extended to include 
another six parameters ($R$, $\sigma_1$, $\sigma_2$, $\alpha_1$,
$\alpha_2$, $\alpha_3$) for topology. 

Topologically non-trivial universes are no longer statistically isotropic.
One must choose an overall orientation of the fundamental domain.
This complicates considerably the evaluation of the likelihood
function for a given topology primarily because one cannot use the 
radical compression
methods \cite{bjkdbbb}. One must work with the full
covariance matrix in real space, $C({\hat \theta},{\hat \theta})$, which
is no longer simply a function of ${\hat \theta}\cdot{\hat \theta}$.
This is a daunting computational task: not  only is one extending
parameter space by another six factors, but one must also marginalize over
all possible orientations in space. Furthermore, what was
once a one-dimensional radial integral over wavenumber becomes
a three dimensional integral which only picks up contributions
from the selected modes on the Fourier grid. If one is to correctly
constrain the properties of the fundamental domain of our
universe, it is essential to develop novel techniques which
can incorporate these various subtleties into likelihood methods.
One should point out, however, the impressive increase in speed of 
the algorithms for performing the various steps
in maximum likelihood estimation. Hopefully this 
means that the challenge of constraining topologically non-trivial
universes  will be met and that it will be possible to do so with the
data which will come from the MAP satellite.
  
{\it Acknowledgments}: 
We thank R.Abrams, J.Levin and J.Silk for discussions.
PGF acknowledges support from the Royal Society.

\tighten

\end{document}